\documentclass[twoside]{article}
\usepackage{amsmath}
\usepackage{amsfonts}
\usepackage{amssymb}
\usepackage{amscd}
\usepackage{graphicx}
\usepackage{color}

\begin{document}
\thispagestyle{plain}

\begin{center}
{\Large \bf \strut
Computational challenges in the relativistic few-nucleon problem
\strut}\\
\vspace{10mm}
{\large \bf 
Wayne Polyzou$^{ab}$
}
\end{center}

\noindent{
\small $^a$\it The University of Iowa, Department of Physics and Astronomy,
Iowa City IA, }\\ 
\noindent{
\small $^a$\it This work supported by the US Department of Energy
undercontract No. DE-FG02-86ER40286}  
\markboth{
W. N. Polyzou}
{
Relativistic few-nucleon problem} 

\begin{abstract}
I discuss computational challenges in the relativistic few-nucleon 
problem and the resolution of some of these challenges.  I also discuss the 
outlook for the future.
\\[\baselineskip] 
\end{abstract}

\section{Introduction}

Studying nuclear physics at distance scales that are potentially
sensitive to sub-nucleon physics requires a relativistic treatment of
the dynamics.  This scale is interesting because QCD is
non-perturabtive at this scale; of particular concern is that it is
not yet known how to compute mathematical error bounds, even for
non-perturbative methods, making the accuracy of calculations based
directly on QCD difficult to assess.  This is also the scale where
transition from meson-nucleon to sub-nucelon degrees of freedom is
poorly understood.

Relativistic quantum mechanics provides a means for studying few-body
problems at this scale.  It provides a quantum mechanical description
of the dynamics of the relevant degrees of freedom consistent with the
exact Poincar\'e symmetry of underlying theory.  Because few-body
models can be solved exactly, comparison of these computations to experiment
provides the direct feedback needed to construct realistic models
based on a given set of degrees of freedom.

Normally the relevant degrees of freedom are the experimental degrees 
of freedom which are the particle spins and momenta that are observed in 
reactions at this scale.  A suitable model Hilbert space is the direct sum
of tensor products of the single-nucleon spaces,
\begin{equation}
{\cal H} = \oplus (\otimes {\cal H}_{m_i\, j_i})
\label{a:1}
\end{equation}
which are irreducible representation spaces for the Poincar\'e group.

Any relativistic model formulated on this space is necessarily
characterized by a unitary representation of the Poincar\'e group
\cite{wigner}
\begin{equation}
U(\Lambda ,a):{\cal H} \to {\cal H} .
\label{a:2}
\end{equation}
The dynamical unitary representation $U(\Lambda, a)$ of the Poincar\'e
group necessarily differs from the natural free-particle
representation, $U_0(\Lambda ,a)$, given by the direct sum of tensor
products free-particle irreducible representations on ${\cal H}$.

The ability to perform local tests of special relativity requires that
the unitary representations of the Poincar\'e group corresponding to
different subsystems be related to $U(\Lambda ,a)$ by  
cluster properties 
\begin{equation}
\lim_{\vert r_{ij} - r_k\vert \to \infty} 
\Vert \left (U(\Lambda ,a) - U_{ij}(\Lambda ,a) \otimes U_k(\Lambda ,a)
\right )
\vert \psi \rangle \Vert  =0 .
\label{a:3}
\end{equation}
The problem of relativistic few-body physics is to construct
mathematical models $U(\Lambda ,a)$ with the above properties that
provide a realistic quantitative and consistent description of few-GeV
scale structure and reactions for few-hadron systems.  This problem is
a natural extension of the corresponding non-relativistic problem; but
the relativistic treatment leads to a number of computational issues
that do not arise in the non-relativistic formulation of the same
problem.

In the non-relativistic case it is useful to work in a frame where the
total momentum $\mathbf{P}$ is zero.  In that frame the Hamiltonian is
replaced by the center of mass Hamiltonian, $h=H-\mathbf{P}^2/2M$,
where $M$ is the Galilean mass of the system.  In the relativistic
case the corresponding operator is the invariant mass operator, which
is the rest energy of the system.  We denote the mass operator by $M$.

The first complication in formulation of a relativistic few-body
dynamics arises because the Hamiltonian appears on the {\it right-hand
  side} of three different commutators.  As a consequence, the
Poincar\'e commutation relations require that at least three of the
Poincar\'e generators have an interaction dependence.  The commutation
relations impose a set of non-linear constraints on these
interactions.  One way to satisfy these constraints is to notice that
all ten generators can be expressed in terms of the two Casimir
operators (mass and spin), four commuting functions of the generators,
and four functions of the generators that are conjugate to the four
commuting functions of the generators.  If interactions are added to
the non-interacting mass operator, keeping these other nine operators
free of interactions, and the ten generators are expressed and
functions of these nine operators and the interacting invariant mass,
the resulting generators will satisfy the Poincar\'e commutation
relations provided the interaction terms commute with these
nine-non-interacting operators.  This is the assumption that defines
the Bakamjian Thomas \cite{bakamjian}method.  These nine commutators 
with the relativistic interaction are the relativistic
equivalent of the nine constraints on the non-relativistic interactions
that result from the requirements that the interactions be translationally
invariant, rotationally invariant, and independent of the total
momentum.

Solving for the mass eigenvalue problem in a suitable irreducible free-particle
basis leads to an explicit dynamical unitary representation of the
Poincar\'e group, $\bar{U}_{ij}(\Lambda ,a)$ on the two particle Hilbert
space.

If this method is applied to the three-nucleon system 
the resulting three-nucleon mass operator \cite{coester} has the from
\begin{equation}
\bar{M} := \bar{M}_{12,3} + \bar{M}_{23,1}+ \bar{M}_{31,2}- 2M_0
\label{a:4}
\end{equation}
\vfill
\begin{equation}
\bar{M}_{ij,k} = M_0 + \bar{V}_{ij} 
\label{a:5}
\end{equation}
\vfill
\begin{equation}
M_0 = \sqrt{\mathbf{q}_k^2 + (\sqrt{\mathbf{k}_{ij}^2 + m_i^2 }
+\sqrt{\mathbf{k}_{ij}^2 + m_j^2 })^2} +
\sqrt{\mathbf{q}_k^2 + m_k^2} 
\label{a:6}
\end{equation}
where the relativistic Jacobi momenta 
\begin{equation}
\mathbf{q}_i := \pmb{\Lambda}(P/M_0)^{-1} p_i  \qquad 
\mathbf{k}_{ij} := \pmb{\Lambda}({q_i + q_j \over m_{0ij}})^{-1} q_i 
\label{a:7}
\end{equation}
are obtained by Lorentz transforming single-particle momenta to the
two and three-body rest frames with non-interacting Lorentz
transformations.  We call these relativistic Jacobi momenta because
the usual Jacobi momenta can be constructed in the same manner by
replacing the Lorentz boost by a Galilean boost.

Because all three of the interactions commute with the same nine functions
of the three-nucleon Poincar\'e generators, 
the interactions can be combined algebraically in the three-nucleon 
mass operator and the result will commute with these same nine operators.
Poincar\'e generators can then be expressed in terms of the interacting 
mass operator and the nine-other three-body kinematic operators.
Again, diagonalizing $\bar{M}$ in a suitable irreducible free-particle 
basis gives a dynamical unitary representation of the Poincar\'e
group, $\bar{U}(\Lambda,a)$, for a system of three interacting 
particles. 

The commutator of the interaction 
with the free spin  operator, $\mathbf{j}_0^2$
\begin{equation}
[\bar{V}_{ij},\mathbf{j}^2_0]=0,
\label{a:8}
\end{equation}
is incompatible with cluster properties of
the three-body Poincar\'e generators.  
The problem is that the {\it relative orbital} angular
momentum, which contributes to the total spin, 
gets modified as a consequence of the interactions.  Here the
failure means that
\begin{equation}
\bar{U}(\Lambda ,a) \rightarrow  \bar{U}_{ij,k}(\Lambda ,a)
\not=
\bar{U}_{ij} \otimes {U}_{k}(\Lambda ,a).
\label{a:9}
\end{equation}
where  $\bar{U}_{ij,k}(\Lambda ,a)$ is obtained from $\bar{U}(\Lambda ,a)$
by turning off the interactions involving particle $k$.
The way that cluster properties fail at the operator level is that 
interactions that should survive in the cluster limit actually vanish.

While cluster properties of $\bar{U}(\Lambda ,a)$ in the sense the 
equation (\ref{a:3}) do not hold, it turns our that the $S$ matrices 
associated with the $2+1$ representations of 
$\bar{U}_{ij,k}(\Lambda ,a)$ and 
$\bar{U}_{ij} \otimes {U}_{k}(\Lambda ,a)$ are identical.

The equivalence of the 2+1 $S$ matrices to the corresponding 
$S$ matrices for the tensor product dynamics 
\begin{equation}
\bar{S}_{ij,k} = \bar{S}_{ij} \otimes I_k 
\label{a:10}
\end{equation}
implies the existence 
\cite{Ekstein} of
an $S$-matrix preserving unitary transformation, $A_{ij,k}$, satisfying
\begin{equation}
A_{ij,k} \bar{U}_{ij,k}(\Lambda ,a) 
A_{ij,k}^{\dagger} = \bar{U}_{ij} \otimes {U}_{k}(\Lambda ,a)  
\label{a:11}
\end{equation}
\begin{equation}
A_{ij,k} \bar{M}_{ij,k} 
A_{ij,k}^{\dagger} = {M}_{ij \otimes k} 
\label{a:12}
\end{equation}
\begin{equation}
A_{ij,k} \mathbf{j}^2_0 
A_{ij,k}^{\dagger} = 
\mathbf{j}^2_{ij\otimes k} 
{ \not=} \mathbf{j}^2_0 . 
\label{a:13}
\end{equation}

Using these unitary operators for each pair of interacting particles 
we construct their Cayely transforms, add the Cayley transforms, and 
inverse Cayley transform the sum of the individual Cayley transforms
to get a new unitary operator $A$\cite{fcwp}: 
\begin{equation}
C_{ij,k} := i (A_{ij,k}-I)(A_{ij,k}+I)^{-1} 
\label{a:14}
\end{equation}
\begin{equation}
C := C_{12,3}+C_{23,1}+C_{31,2} 
\label{a:15}
\end{equation}
\begin{equation}
A:=(I - i C)(I+i C)^{-1} \qquad A \to A_{ij,k} \to I .
\label{a:16}
\end{equation}
The resulting transformation $A$ is an $S$-matrix preserving 
unitary transformation.  Using it to transform 
$\bar{U}(\Lambda ,a)$ gives a new unitary representation\cite{sokolov} of the 
Poincar\'e group 
\[
U(\Lambda ,a) := A^{\dagger} \bar{U}(\Lambda ,a) A
\]
satisfying cluster properties (\ref{a:3}) of the unitary representation of the 
Poincar\'e group 
\begin{equation}
U(\Lambda ,a) \to \bar{U}_{ij}(\Lambda ,a)
\otimes {U}_{k}(\Lambda ,a).
\label{a:17}
\end{equation}
The Poincar\'e generators for this representation include sums of the 
different pairwise interactions.  The operators $A$ and $A_{ij,k}$
also generate additional three-nucleon forces that are needed 
to satisfy the commutation relations.  These three-nucleon forces are
different from standard three-nucleon forces because 
they are frame-dependent {\it and} are explicit functions of the 
underlying two-nucleon forces.

The resulting invariant mass operator has the form
\begin{equation}
M = A (\sum A^{\dagger}_{ij,k} {M}_{ij \otimes k} A_{ij,k}- 2M_0 )A^{\dagger}
= A \bar{M} A^{\dagger}.
\label{a:18}
\end{equation}
The important property is that because $A$ is S-matrix preserving it
means the $\bar{M}$ leads to the same $S$ matrix as $M$, so even
though the representation $\bar{U}(\Lambda ,a)$ fails to satisfy
(\ref{a:3}), it has the same $S$ matrix as the model satisfying
cluster properties.  This means that for scattering and bound state 
calculations, {\it it is sufficient to solve the Faddeev equations for 
$\bar{M}$}. 

This avoids that complications of computing the additional
three-nucleon interaction that appears in $M$ in the three-body case,
however it is important to remark that this equivalence does not
extend to the four-nucleon case unless the corresponding generated
three-body interactions appear in the four-body mass operator.  We
also remark the two-body interactions $\bar{V}_{\gamma}$ are really
three-body operators due to the role of the spectator momentum - one
can think of them as frame-dependent two-body interactions.

The next set of complications are more technical.  In order to formulate 
relativistic Faddeev equations for the dynamics given by the 
mass operator $\bar{M}$ we define the operators
\begin{equation}
\bar{M}=M_0 +\bar{V} \qquad \bar{V}= \sum_{\alpha} \bar{V}_{\alpha}  
\qquad \alpha \in \{ (12,3),(23,1),(31,2) \}
\label{a:19}
\end{equation}
\begin{equation}
\bar{V}_\alpha = \bar{M}_\alpha -M_0    \qquad \bar{V}^{\alpha} = \bar{M}-
\bar{M}_{\alpha} .
\label{a:20}
\end{equation}
Using time-dependent methods \cite{bkwp} 
it is possible to show that the $S$ matrix can
be expressed in terms of the following relativistic transition operator
\begin{equation}
\bar{T}^{\alpha \beta}(m) := 
\bar{V}^\beta + \bar{V}^{\alpha}(m-\bar{M}+i 0^+)^{-1} \bar{V}^{\beta} 
\label{a:21}
\end{equation}
\begin{equation}
\langle a_0 \vert S^{\alpha \beta}\vert  b_0 \rangle  = 
\langle a_0 \vert b_0\rangle  - 2 \pi i 
\langle a_0 \vert \delta (m_a - m_b)
 { \bar{T}^{\alpha\beta} (m_a+i0^+)}\vert b_0 \rangle .
\label{a:22}
\end{equation}
The different components of $\bar{T}^{\alpha \beta}(m)$ satisfy 
the relativistic Faddeev equation
\begin{equation}
\bar{T}^{\alpha\beta}(z) = \bar{V}^\beta + \sum_{\gamma \not= \alpha} 
\bar{T}_\gamma (z-M_0)^{-1} \bar{T}^{\gamma\beta}(z). 
\label{a:23}
\end{equation}
The input to (\ref{a:23}) equation is the $2+1$ transition operators 
\begin{equation}
\bar{T}_\gamma (z) = \bar{V}_\gamma +  
\bar{V}_\gamma (z-M_0)^{-1} \bar{T}_\gamma (z).
\label{a:24}
\end{equation}
As in the non-relativistic case the Faddeev equation can be solved with 
mathematically controlled errors because the iterated kernel is 
compact and can be uniformly approximated by a finite dimensional matrix:
\begin{equation}
\bar{T}(z) = \bar{D}(z) + \bar{K}(z)\bar{T}(z)  
\qquad \bar{K}(z)^2 \quad \mbox{\bf compact}
\label{a:25}
\end{equation}
\begin{equation}
\bar{T}(z) = (I- \bar{K}(z)^2)^{-1} (\bar{D}(z) + \bar{K}(z)\bar{T}(z)).
\label{a:26}
\end{equation}

The first technical problem is to construct realistic two-nucleon
interactions.  Repeating what was done for the non-relativistic
problem, by carefully fitting models to two-nucleon phase shifts, can
also be done in the relativistic case, but because both the relativistic and
non-relativistic interactions are fit to the same data, refitting is
not necessary.  The trick was first given by Coester, Pieper and
Serduke\cite{cps}.

The mass operator in the Bakamjian-Thomas representation has the form
\begin{equation}
\bar{M} := M_0 + \bar{V}_{12} + \bar{V}_{23}+ \bar{V}_{31}
\label{a:27}
\end{equation}
where 
\[
\bar{V}_{ij} :=
\]
\[
\sqrt{\mathbf{q}_k^2 + (\sqrt{{\mathbf{k}_{ij}^2} + m_i^2 + {2 \mu_{ij}
v_{nr\,ij}}}
+\sqrt{{\mathbf{k}_{ij}^2} + m_j^2 + 
{ \mu_{ij}v_{nr\,ij}}})^2} -
\]
\[
\sqrt{\mathbf{q}_k^2 + (\sqrt{\mathbf{k}_{ij}^2 + m_i^2}
+\sqrt{\mathbf{k}_{ij}^2 + m_j^2 })^2} 
\]
\begin{equation}
\bar{M}_{ij,k} = M_{ij,k}(h_{nr\,ij})
\label{a:28}
\end{equation}
and $\mu_{ij}$ is the two-nucleon reduced mass.
The important property of this interaction is the corresponding $2+1$ mass 
operator is a function of the non-relativistic nucleon-nucleon rest 
Hamiltonian, $h_{ij}= H_{ij}- {(\mathbf{p}_{i}+\mathbf{p}_j)^2
\over 2(m_i+m_j)}$.
This means that the $S$ matrix in both the relativistic and non-relativistic
models have the same internal wave functions and phase shifts as a 
function of the center of mass momentum $\mathbf{k}$:
\begin{equation}
\langle \mathbf{p}, \mathbf{q}_r, \mathbf{k}_r \vert S_{ij,k\, r} \vert
\mathbf{p}', \mathbf{q}_r', \mathbf{k}_r' \rangle = 
\delta (\mathbf{p}- \mathbf{p}' )\delta (\mathbf{q}_r-\mathbf{q}_r' )
{\langle \mathbf{k}_r \vert s_{ij} \vert \mathbf{k}_r' \rangle}
\label{a:29}
\end{equation}
\begin{equation}
\langle \mathbf{p}, \mathbf{q}_{nr}, \mathbf{k}_{nr} \vert S_{ij,k\, nr}
\vert 
\mathbf{p}', \mathbf{q}_{nr}', \mathbf{k}_{nr}' \rangle =
\delta (\mathbf{p}- \mathbf{p}' )\delta (\mathbf{q}_{nr}-\mathbf{q}_{nr}' )
{\langle \mathbf{k}_{nr} \vert s_{ij} \vert \mathbf{k}_{nr}' \rangle}.
\label{a:30}
\end{equation}
In order to take advantage of this relationship we recall that the
two-body input to the relativistic Faddeev equation can be expressed
in the following ways
\begin{equation}
\langle \mathbf{p}, \mathbf{q}_r, \mathbf{k}_r\vert
\bar{T}_{\alpha} \vert \mathbf{p}', \mathbf{q}'_r, \mathbf{k}'_r \rangle =
\langle \mathbf{p}, \mathbf{q}_r, \mathbf{k}_r \vert
\bar{V}_{\alpha} \vert \mathbf{p}', \mathbf{q}'_r, \mathbf{k}_r^{\prime -} \rangle =  
\langle \mathbf{p}, \mathbf{q}_r, \mathbf{k}_r \vert
(\bar{M}_{\alpha}-M_0) \vert \mathbf{p}', \mathbf{q}'_r, \mathbf{k}_r^{\prime -} \rangle .
\label{a:31}
\end{equation}
Since for the above choice of interaction the internal 
relativistic and non-relativistic wave functions
are identical we get the identifications
\begin{equation}
\langle \mathbf{k}  \vert \mathbf{k}^{\prime -}_{nr}  \rangle =
\langle \mathbf{k}  \vert \mathbf{k}^{\prime -}_r \rangle .
\label{a:31a}
\end{equation}
Using this it follows that the Faddeev kernel can be written as 
\[
\langle \mathbf{q}_\alpha, \mathbf{k}_{\alpha} \vert
T_{\alpha}(z) (z-\bar{M}_{0})^{-1}  \vert \mathbf{q}_\alpha', \mathbf{k}_{\alpha}' \rangle = 
\]
\[
\delta(\mathbf{q}_{\alpha}-\mathbf{q}'_{\alpha}) 
{m_{0\alpha}(\mathbf{k}) + m_{0\alpha}(\mathbf{k}') \over 
( \sqrt{\mathbf{q}_{\alpha}^2 + m_{0\alpha}^2 (\mathbf{k}_{\alpha})} +
\sqrt{\mathbf{q}_{\alpha}^2 + m_{0\alpha}^2(\mathbf{k}_{\alpha}')}
)} \times
\]
\begin{equation}
{\langle \mathbf{k}_{\alpha} 
\vert t_{r}(z) 
\vert \mathbf{k}'_{\alpha} \rangle} 
{1 \over M_{0}(\mathbf{q}_{\alpha}, \mathbf{k}_{\alpha} )- 
M_{0}(\mathbf{q}_{\alpha}, \mathbf{k}'_{\alpha} )+i0^+}
\label{a:32}
\end{equation}
where 
\begin{equation}
m_{0\alpha}(\mathbf{k}_{\alpha}) :=
\sqrt{\mathbf{k}_{\alpha}^2 + m_i^2} + \sqrt{\mathbf{k}_{\alpha}^2 + m_j^2} 
\label{a:33}
\end{equation}
and
\[
z= M_0 (\mathbf{q}_\alpha \mathbf{k}_\alpha)+ i0^+ ,
\]
\[
\langle \mathbf{k}_{\alpha} 
\vert t_{r}(z) 
\vert \mathbf{k}'_{\alpha} \rangle =
\]
\[
({2 \mu \over \sqrt{\mathbf{k}^2_{\alpha} + m_i^2}+ 
\sqrt{\mathbf{k}^{\prime 2}_{\alpha} + m_j^2}}  + 
{2 \mu \over \sqrt{\mathbf{k}_{\alpha}^2 + m_i^2}+ 
\sqrt{\mathbf{k}_{\alpha}^{\prime 2} + m_j^2}}) 
\times 
\]
\begin{equation}
{\langle \mathbf{k}_{\alpha} 
\vert t_{nr}(\mathbf{k}_{\alpha}^2/2\mu+i0^+)
\vert \mathbf{k}'_{\alpha} \rangle} .
\label{a:34}
\end{equation}
These relations express the Faddeev kernel in terms of the
{\it non-relativistic transition} matrix elements.  The identity of the 
wave functions, which was used to
derive the result, is limited to the case that the transition matrix
elements are half-on shell.  This relations does not extend to 
the off-shell transition matrix elements which appear in the
Faddeev kernel.

The fully off-shell two-body $\bar{T}_{\alpha}(z)$ embedded in the 
three-nucleon Hilbert space can be computed by solving the first 
resolvent equation\cite{charlotte}:
\begin{equation}
\bar{T}_{\alpha}(z)= \bar{T}_{\alpha}(z') + \bar{T}_{\alpha}(z) {z'-z \over (z-M_0)(z'-M_0)} 
\bar{T}_{\alpha}(z').  
\label{a:35}
\end{equation}

Finally we note that while it is natural to use variables to 
label two-nucleon interactions to be associated with the 
two-nucleon rest frames, with theses variables the permutation 
operators involve Wigner rotations.  The Wigner rotations 
can be removed from the 
permutation operators by expressing everything in terms of 
variables associated with the three-nucleon rest frame.
In this representation the Wigner rotations appear in the elementary 
nucleon-nucleon interactions:
\[
\langle \mathbf{q}_i, \mu_i, \mathbf{q}_j, \mu_j \vert t_{r}(z) 
\vert \mathbf{q}_i', 
\mu_i', \mathbf{q}_j' , \mu_j' \rangle = 
\]
\[
\left ( {\omega_i (\mathbf{q}_{i})+\omega_j(\mathbf{q}_{j}) 
\over \omega_i (\mathbf{k}_{ij})+\omega_j(\mathbf{k}_{ji}) }
{\omega_i (\mathbf{k}_{ij}) \over  
\omega_i (\mathbf{q}_{i})} 
{\omega_j(\mathbf{k}_{ji})\over 
\omega_j(\mathbf{q}_{j}) }
\right )^{1/2} \times
\]
\[
\sum 
D^{j_i}_{\mu_i \nu_i}
[R_{wc}(B_c(q_{ij}),k_{ij})]
D^{j_j}_{\mu_j \nu_j}
[R_{wc}(B_c(q_{ij}),k_{ji})] \times
\]
\[
{\langle \mathbf{k}_{ij}, \nu_i, \nu_j 
\vert t_{r}(z) \vert \mathbf{k}_{ij}', 
\nu_i', \nu_j' \rangle }\times
\]
\[
D^{j_i}_{\nu_i' \mu_i'}
[R_{wc}(B^{-1}_c(q_{ij}),q_{i})]
D^{j_j}_{\nu_j' \mu_j'}
[R_{wc}(B^{-1}_c(q_{ij}),q_{j})] \times
\]
\begin{equation}
\left ( {\omega_i (\mathbf{q}'_{i})+\omega_j(\mathbf{q}'_{j}) 
\over \omega_i (\mathbf{k}'_{ij})+\omega_j(\mathbf{k}'_{ji}) }
{\omega_i (\mathbf{k}'_{ij}) \over  
\omega_i (\mathbf{q}'_{i})} 
{\omega_j(\mathbf{k}'_{ji})\over 
\omega_j(\mathbf{q}'_{j}) }
\right )^{1/2}.
\label{a:36}
\end{equation}

The final technical challenge is that at the few-hundred MeV scale
partial-wave projections begin to loose their underlying advantage.
This is in part because the transition operator is a relatively smooth
operator, so there are necessarily a lot of cancellations involved in
the partial wave expansions, especially at large angles.  As a
practical matter double precision three-nucleon calculations based on
partial wave methods are limited to about 300 MeV.  Direct integration
calculations are stable over a wider range of energies\cite{charlotte},
extending to the few-GeV scale.

The final computational challenge is that the natural input to
direct-interaction three-nucleon calculations is a momentum-space
interaction in operator form.  One of the few realistic interactions
in operator form is the Argonne V18 interaction which is given in a
configuration-space representation.

It has been Fourier transformed \cite{veerasamy} in an operator form.  
The resulting interaction can be expanded in terms of 24 spin-isospin 
operators.

It is possible to reduce the number of required operators using symmetry 
properties.  The most general nucleon-nucleon interactions can be expanded 
in terms of the following spin operators:
\begin{equation}
\langle \mathbf{k} \vert v_{nr} \vert
\mathbf{k}' \rangle = \sum V_n  
W_n 
\label{a:37}
\end{equation}

\begin{equation}
W_1 := I
\label{a:38}
\end{equation}

\begin{equation}
W_2 :=\mathbf{j}_1 \cdot \mathbf{j}_2
\label{a:39}
\end{equation}

\begin{equation}
W_3 := (\mathbf{j}_1 \cdot \hat{\mathbf{K}}) \otimes
(\mathbf{j}_2 \cdot \hat{\mathbf{K}}) 
\label{a:40}
\end{equation}

\begin{equation}
W_4 := (\mathbf{j}_1 \cdot \hat{\mathbf{Q}})\otimes
(\mathbf{j}_2 \cdot \hat{\mathbf{Q}}) 
\label{a:41}
\end{equation}

\begin{equation}
W_5 :=  (\mathbf{j}_1 \cdot \hat{\mathbf{N}})
\otimes I_2 + I_1 \otimes 
(\mathbf{j}_2 \cdot \hat{\mathbf{N}}) 
\label{a:42}
\end{equation}
\begin{equation}
W_6 := (\mathbf{j}_1 \cdot \hat{\mathbf{K}})\otimes
(\mathbf{j}_2 \cdot \hat{\mathbf{Q}}) +
(\mathbf{j}_1 \cdot \hat{\mathbf{Q}})\otimes
(\mathbf{j}_2 \cdot \hat{\mathbf{K}})
\label{a:43}
\end{equation}
where
\begin{equation}
\mathbf{K} :=\mathbf{k}'-\mathbf{k}
\qquad
\mathbf{Q} :=\mathbf{k}'+\mathbf{k}
\qquad
\mathbf{N} :=\mathbf{k}'\times \mathbf{k} .
\label{a:44}
\end{equation}
The coefficients of these operator expansions are simply related to
the Wolfenstein parameters\cite{walter}, which facilitates the
computation of spin observables.  The remaining computational
difficulty is related to the observation that there are five independent 
operators on shell, and one more off shell.

Numerical instabilities can arise when the independent on-shell and
off-shell operators are not simply related\cite{veerasamy2}.  For the
choice above five of the off-shell operators become the five 
on-shell operators in the on-shell limit.

The last dynamical consideration is the computation of current matrix
elements, which are needed to study few-nucleon systems with 
few-GeV scale hadronic probes.  The important observation is that 
any change of representation of the Poincar\'e generators 
requires a corresponding change of representation of the current operator
in order leave the physical observables unchanged.   In principle one 
expects both the strong dynamics and electromagnetic current to 
satisfy cluster properties.  This suggest that using currents that 
have well-behaved cluster expansions should not be used in 
Bakamjian-Thomas representation of the dynamics.  In general one expects that 
one must first transform either the current operator or the dynamics 
with an operator like (\ref{a:16}): 
\begin{equation}
\langle \Psi_f \vert J^{\mu} (0) \vert \Psi_i \rangle =
\langle \bar{\Psi}_f \vert 
A^{\dagger}J^{\mu} (0)A \vert \bar{\Psi}_i \rangle \approx 
\langle \bar{\Psi}_f \vert 
J^{\mu} (0) \vert \bar{\Psi}_i \rangle .    
\label{a:45}
\end{equation}
When $A$ is close to the identity, which appears to be the case for
nuclear physics scales (\cite{bkwp}), this operator can be ignored,
resulting in a significant increase in computational efficiency.

As a result of these various simplifications and tricks it has been
possible to perform three-nucleon calculations with realistic
interactions \cite{walter-witala}.  Figure 1 show the differential
cross section for $p-d$ elastic scattering for relativistic and
non-relativistic three-nucleon models with realistic two- (CD Bonn)
and three-nucleon (TM99) interactions.  The calculations show that for
elastic scattering the relativistic effects are small, except at back
angles, where there is some enhancement due to relativity for the 250
MeV curves.  Comparison of these calculations with measurements from
\cite{seguchi} shows that there is missing physics that is not
explained by the combination of the $TM99$ three-nucleon force and
relativity.  Elastic spin observables at these also show a weak 
dependence on relativistic effects.  This is in part the comparison that
we show is only sensitive to the difference in how the two-nucleon 
subsystem is embedded in the three nucleon system.  Breakup calculations,
on the other hand, exhibit strong relativistic effects in certain 
observables. 
The calucaltions in
figure 2 \cite{ting} provide a beautiful illustration of some of these effects.
These calculations were at a much higher energy than the calculations
of figure 1 however they only use spin-independent Malfliet-Tjon interaction.
The figure show the fivefold differential cross section where the scattered
protons emerge symmetric at different angles relative to the beam 
line. These are plotted against the energy of one of the scattered protons.
This figure shows a dramatic crossing of the non-relativistic and relativistic
results as the angle is changed.  The data is from \cite{punjab}.

In this manuscript we have discussed many of the complications
involved in making realistic relativistic three-nucleon calculations.
We have discussed tricks that make realistic calculations possible at
relativistic energies.  The calculations suggest that the relativistic
effects are small for nucleon-meson degrees of freedom, except in
certain areas of breakup phase space, however realistic relativistic
calculations have not been performed at the few-GeV scale.  The
discrepancy of the calculated large-angle elastic scattering cross
section with data suggests some missing short distance physics in the
three-nucleon forces.

We anticipate that relativistic few-body methods will be an important
tool for understanding physics at scales between the Chiral
perturbation theory and pertrubative QCD scales.  Modern computers
have made realistic few-GeV scale few-body calculations feasible.  The
approach that we advocate, using models with the dominant degrees of
freedom and symmetries is similar to the approach used in condensed
mater physics.  It is far easier than attempting to get mathematical
convergent approximations of QCD at the Few GeV scale.

%

\begin{figure}
\centerline{\includegraphics[width=0.8\textwidth]{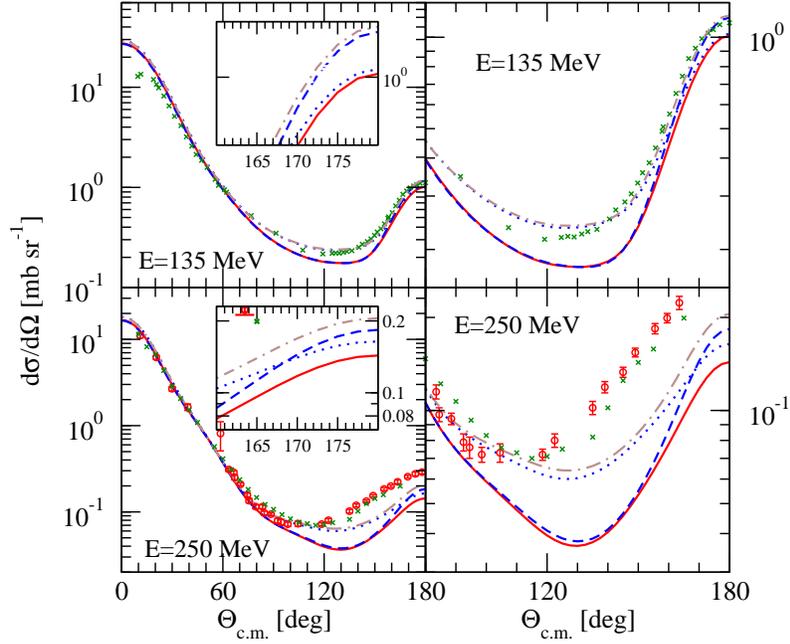}}
\caption{Relativistic effects in elastic p-d scattering}
\label{fig1}      
\end{figure}

\begin{figure}
\centerline{\includegraphics[width=0.8\textwidth]{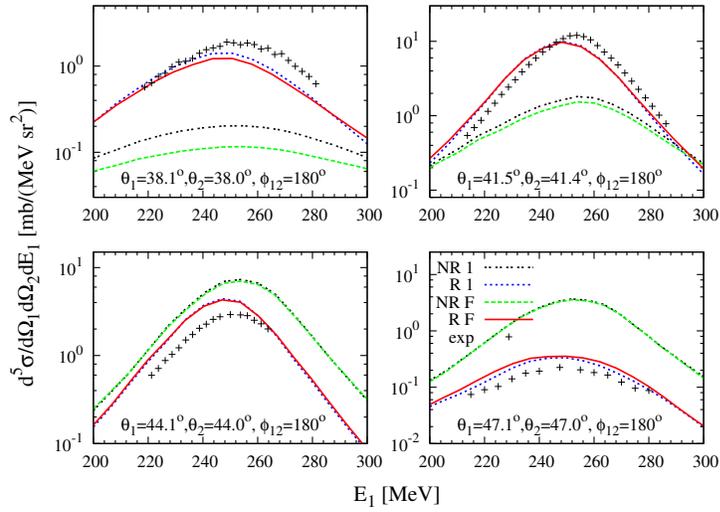}}
\caption{Relativistic effects in n-d breakup reations}
\label{fig2}      
\end{figure}

This research supported by the U.S. Department of Energy Office of Science

\end{document}